\documentclass{article} 
\usepackage{iclr2020_conference,times}

\usepackage{hyperref}
\usepackage{url}
\usepackage{graphicx}

\title{Surrogate sea ice model enables \\efficient tuning}

\author{Kelly~Kochanski 
Department of Geological Sciences\\
University of Colorado Boulder\\
Boulder, CO, USA \\
\texttt{kelly.kochanski@colorado.edu} \\
\And
Ivana~Cvijanovic\\
Centro Nacional de Supercomputaci\'on\\
Barcelona, Spain \\
\AND
Donald~Lucas \\
National Atmospheric Release Advisory Center \\
Lawrence Livermore National Laboratory\\
Livermore, CA, USA \\
}

\iclrfinalcopy 
\begin{document}

\maketitle

\begin{abstract}
Predicting changes in sea ice cover is critical for shipping, ecosystem monitoring, and climate modeling.
Current sea ice models, however, predict more ice than is observed in the Arctic, and less in the Antarctic.
Improving the fit of these physics-based models to observations is challenging because the models are expensive to run, and therefore expensive to optimise.
Here, we construct a machine learning surrogate that emulates the effect of changing model physics on forecasts of sea ice area from the Los Alamos Sea Ice Model (CICE).
We use the surrogate model to investigate the sensitivity of CICE to changes in the parameters governing: ice's ridging and albedo, snow's albedo, aging, and thermal conductivity, the effect of meltwater on albedo, and the effect of ponds on albedo.
We find that the CICE's sensitivity to these model parameters differs between hemispheres. 
We propose that future sea ice modelers separate the snow conductivity and snow grain size distributions on a seasonal and inter-hemispheric basis, and we recommend optimal values of these parameters.
This will make it possible to make models that fit observations of both Arctic and Antarctic sea ice more closely.
These results demonstrate that important aspects of the behaviour of a leading sea ice model can be captured by a relatively simple support vector regression surrogate model, and that this surrogate dramatically increases the ease of tuning the full simulation.
\end{abstract}

\section{Motivation}
\label{sec:motivation}
Understanding future sea ice extents is critical for naval and shipping strategies, 
conserving polar ecosystems, and modelling Earth's energy balance and climate.
Current models of sea ice, however, consistently predict more extensive Arctic sea ice than is observed \citep{Stroeve2007,Perovich2019},
and less extensive Antarctic sea ice \citep{Eisenman2011,Stroeve2018}.
This misfit may be caused by missing or mis-calibrated model physics.


Sea ice models are designed to capture the physical processes that shape snow and ice 
on 10--30\,km length scales,
with smaller-scale processes represented through spatially averaged parameters \citep{Hunke2017}.
The values of these parameters are tuned to observational data \citep{Kim2006},
but the tuning process is challenging, as each run of a global sea ice model can take days to weeks.
Here, we develop a machine learning surrogate of the leading sea ice model, CICE \citep{Hunke2017},
use the surrogate model to efficiently evaluate the model's response to changing parameters,
demonstrate that the Arctic and Antarctic must be tuned separately,
and provide recommendations for reducing the current model-data misfit.

\section{Methods}
Our model is designed to enable accelerated analysis and tuning of the Los Alamos Sea Ice Model, CICE. This is a global sea ice model used for operational ice forecasts, and its thermodynamic core is used in many sea ice and climate models worldwide. Thus, improving the accuracy of CICE is an effective route towards improving many widely-used sea ice forecasts. 
Although some authors have begun to use machine learning to predict the properties of current sea ice \citep{Lee2016} machine learning is not yet integrated into operational forecasts.

Long-term sea ice trends are well-predicted by physics-based models. The cost of these models is high enough to make tuning inconvenient, but not so high as to warrant replacing the models entirely with ML systems. Therefore, we focus on making improvements to a widely-used model, rather than beginning from the ground up with a pure machine learning pipeline like \citet{Chi2017}. This approach allows us to leverage the speed of an ML emulator in the context where it is most effective, while retaining the predictive power of the physics-based model, and working within existing ice forecast infrastructure.

We sought a surrogate model that could capture the effect of selected model physics (\textsection\ref{sec:param selection}) on CICE ice cover forecasts in realistic past atmospheric conditions (\textsection\ref{gen_inst}), with the goal of identifying strategies that simultaneously
(1)~decrease the model's prediction of Arctic ice cover, 
(2)~increase the prediction of Antarctic ice cover, and 
(3)~improve the realism of the model physics.

\subsection{Selection of target parameters}
\label{sec:param selection}
We selected seven CICE parameters that represent ice and snow processes with strong influences on total ice growth and area. These are listed in Table~\ref{tab:parameters varied}, along with their default values in CICE 4.0. The functions of these parameters are described fully in \citet{Hunke2015}. The parameter selection was guided by sensitivity studies of CICE \citep{Urrego-Blanco2016,Blazey2013}, which brought attention to the high degrees of uncertainty caused by parameterizations of snow and ice albedo  (reflectivity), snow thermal conductivity, and snow aging.

\begin{table}[]
    \caption{Parameters varied in CICE perturbed parameter ensemble.}
        \centering
\begin{tabular}{l l l l}
     \textbf{Parameter} &  & \textbf{Range}    & \textbf{Default}\\
     \hline
     Snow thermal conductivity          & \verb|ksno|          & 0.10--0.35    & 0.30\\
     Snow aging albedo parameter       & \verb|r_snw|        & -1.9--1.9     & 1.5\\
     Albedo adjustment from ice on snow     & \verb|r_ice|        & -1.9--1.9     & 0.0\\
     Albedo adjustment from ponds on snow    & \verb|r_pnd|        & -1.9--1.9     & 0.0\\
     Snow melt maximum radius       & \verb|rsnw_melt_in|           & 500--2000     & 1500\\
     Melting onset temperature, $^oC$   & \verb|dT_mlt_in|    & 0.10--1.80    & 1.50\\
     e-folding scale of ice ridges      & \verb|mu_rdg|       & 3.0--5.0          & 4\\
\end{tabular}
    \label{tab:parameters varied}
\end{table}

\subsection{Data generation}
\label{gen_inst}
From these parameters, we generated a perturbed parameter ensemble of 72 CICE simulations. Parameter combinations were selected using a Latin hypercube distribution, produced by the LLNL Uncertainty Quantification Pipeline \citep{Brandon2011,Tannahill2011} within the ranges specified in Table~\ref{tab:parameters varied}.

To control the cost of generating this training data, we ran the ensemble with a repeated year 2000 atmospheric and ocean forcing.
This is less realistic than a fully-coupled atmosphere that responds to changes in the sea ice, but tens of times faster.
The atmosphere was modelled by the Community Atmosphere Model (CAM4, grid f19); the ocean as a slab ocean; the sea ice model used a displaced pole grid (gx1v6); and the land surfaces using the Community Land Model \citep{Oleson2010}. The simulations ran for 40 years, the first 30 of which were used to spin up the model. To average out the internal variability of the climate, we averaged the last 10 years of the runs together.

\subsection{Surrogate model design}
The surrogate model predicts ice area and extent as a function of seven
CICE model parameters (\verb|r_snw|, \verb|r_pnd|, \verb|r_ice|, \verb|dt_melt_in|, \verb|rsnw_melt_in|, \verb|ksno| and \verb|mu_rdg|) and a specific query (ice area or extent, in either the northern or southern hemisphere, in a specific month).
The surrogate model is a support vector regression \citep{Smola2004}, implemented in Python using scikit-learn \citep{Pedregosa2011}.

\subsection{Testing and validation}
\begin{figure}[hbtp]
    \centering
    \includegraphics[width=.45\textwidth]{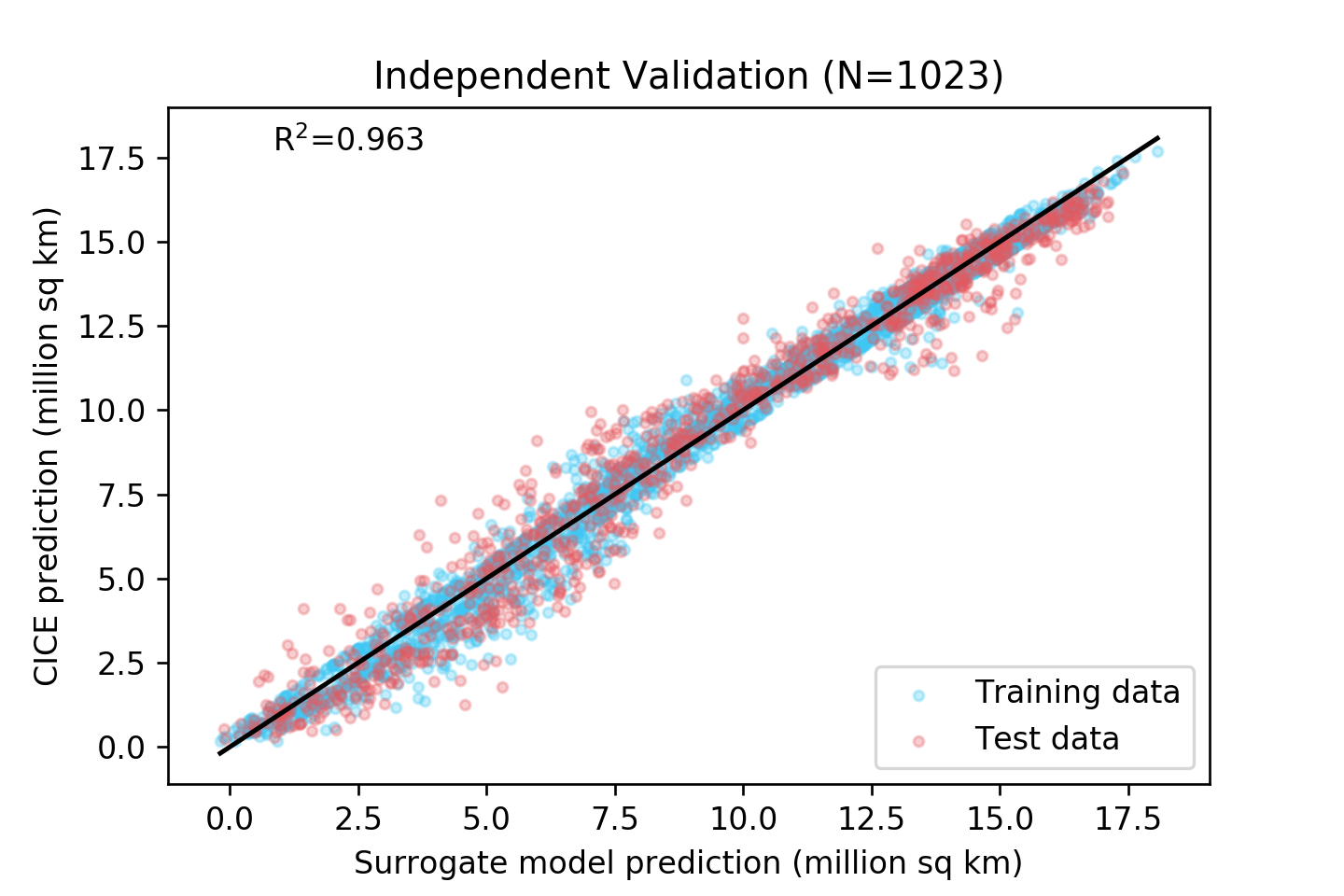}
    \includegraphics[width=.4\textwidth]{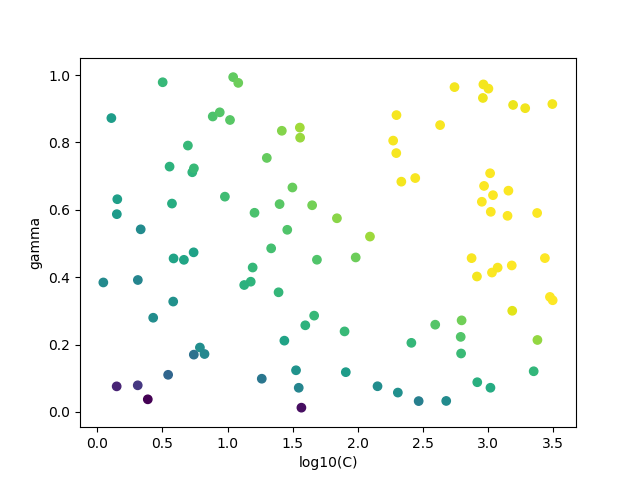}
    \caption{Surrogate model tuning and validation. Left: Fit between tuned surrogate model output and CICE output. Right: Variation in surrogate model fit as a function of hyperparameters $C$ and $\gamma$ for fixed $\nu=0.45$. Lighter colors represent better fits.}
    \label{fig:svr fit}
\end{figure}

The model was trained and tested on 3360 data points (each representing ice area or extent, in a given month, in a given hemisphere, for one of the 72 members of the perturbed parameter ensemble).
We split test and training data randomly at a ratio of 3:7. 
The model is not particularly sensitive to the test/training data split: a fifteen-fold cross-validation with random splits returned R$^2$ goodness of fit scores varying between 0.952--0.968. The model used in the final study (Fig.~\ref{fig:svr fit}) has a typical value of R$^2$=0.963.

Our SVR model's behavior is governed by a set of hyperparameters: $C$, $\gamma$, and $\nu$. We tuned the model by searching this hyperparameter space for the best fit (Fig.~\ref{fig:svr fit}b). 
We settled on $C=318$, $\gamma=0.868$, and $\nu=0.349$ to balance surrogate model speed (which decreases with $C$) and accuracy, and used these values for the remainder of the study.

\subsection{Bayesian calculation of model-data fit}
We compare the CICE model results to observed data from the NSIDC sea ice index \citep{Fetterer2017}. Both modelled and observed data consist of monthly averages of ice area and extent in 2000. The difference between our model and the observations is its \emph{misfit}. Misfit approaches zero as the model--data fit becomes perfect (or, as the likelihood that the observed data was produced by that model becomes high), is negative if the model predicts less ice than was observed, and is positive if the model predicts more ice than was observed. The misfit is given by
\begin{equation}
    \mathrm{misfit}(\vec{\theta}) = \left(M_0\right)\left(\vec{x}(\vec{\theta}) - \vec{y}\right)\left({\mathrm{cov}}^{-1}\right)\left(\vec{x}(\vec{\theta}) - \vec{y}\right)
\end{equation}
where $\vec{\theta}$ is a set of parameters, $\vec{y}$ is observational data, $\vec{x}$ is model data, $\mathrm{{cov}}^{-1}$ is the inverse covariance matrix for the observational data, and $M_0$ is a normalization constant such that the misfit for the default parameter settings (Table~\ref{tab:parameters varied}) has magnitude 1.

The magnitude of the $\mathrm{misfit}$ is inversely related to the posterior probability, $p(\vec{\theta}|\vec{y})$, that known observations $\vec{y}$ could be produced by a model with parameters $\vec{\theta}$, assuming a uniform prior probability for those parameters. Thus, the set of parameters that minimizes the absolute value of the misfit has the highest likelihood of reproducing the observational data.

\section{Results}

\begin{figure}[hbtp]
    \centering
    \includegraphics[width=.45\textwidth]{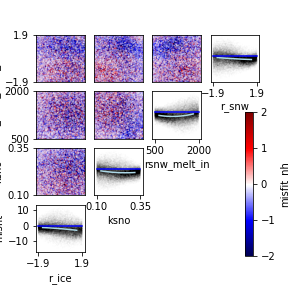}
    \includegraphics[width=.45\textwidth]{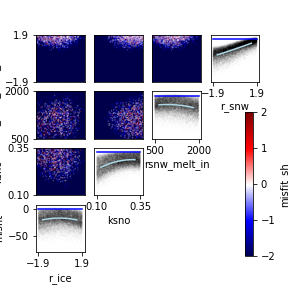}
    \caption{Likelihood that different sea ice model (CICE) parameterizations would have reproduced observed ice area and extent in year 2000.
 Left: probability of producing Arctic observations. 
 Right: probability of producing Antarctic observations.
 Colored plots are bivariate distributions, white background plots are univariate distributions with trend lines. Blue indicates that the model produces less ice than is observed whereas red indicates that the model produces excess ice; white indicates good agreement.}
    \label{fig:statistics}
\end{figure}

The surrogate model is dramatically faster than CICE: each data point for ice area and extent is generated in $\sim$5 ms, versus $\sim$96 CPU hours for each of the model runs described above --- about $7\times10^7$ times faster.
This made it possible to explore the model's behavior through a Monte-Carlo study of 600,000 parameter combinations.
The resulting model--data misfits are shown for selected parameters in Fig.~\ref{fig:statistics}.


These results reveal that Arctic and Antarctic sea ice respond differently to changes in model parameters. 
Specifically, the parameter range we explored contains many good fits to the Arctic data (white dots in Fig.~\ref{fig:statistics} left), but most of those parameter combinations produce far too little ice (blue) in the Antarctic (Fig.~\ref{fig:statistics} right). 
Moreover, the two hemispheres have opposite sensitivities to some parameters: increasing \verb|r_snw| or \verb|ksno| strongly increases the ice cover in the Antarctic, but not in the Arctic.

\section{Implications for sea ice model design}
\label{others}
We find that the effects of sea ice parameters on predicted ice cover differ between the Arctic and Antarctic.
We recommend that sea ice model developers treat parameters such as \verb|ksno| and \verb|r_snw| not as global parameters, but as dynamic properties that may change in space or time. 
Specifically, as a simple, high-level fix, we recommend that developers use a lower value of the thermal conductivity of snow, \verb|ksno|, in the Antarctic than the Arctic, and a higher value of the snow aging parameter \verb|r_snw|.
Implementing this change could quickly create models that better fit observations of Antarctic sea ice, and would add physical realism: both the thermal conductivity and the effect of aging on snow differ between the hemispheres \citep{Sturm2002,Massom2001}.
 
In future work, we intend to validate this recommendation by implementing it within CICE, and exploring the effect of improved sea ice model parameters on the global energy balance.
We have also limited this discussion to variations of the largest possible scale. In future work, we intend to explore the effect of these model parameters on regional and seasonal sea ice variability, as well as the inter-hemisphere variability explored here. 
Field studies have shown that the physics of sea ice is variable on many scales, ranging from regional differences in snowfall and wind speed \citep{Holland2012}, to meter-scale variability in the form of ice leads or snow dunes \citep{Popovic18,Kochanski2018}.

This study shows that the behaviours of the leading sea ice model CICE can be captured by a relatively simple support vector regression surrogate model and that this surrogate dramatically increases the ease of exploring the behaviours of the full simulation. This facilitates better decisions about the structure of the physics-based model.


\bibliography{seaice}
\bibliographystyle{iclr2020_conference}

\end{document}